\documentstyle[12pt,epsfig]{article}
\newcommand{\dafne}{\mbox{DA$\Phi$NE}}
\newcommand{\pio}{\mbox{$\pi^{o}$}}
\newcommand{\pim}{\mbox{$\pi^{-}$}}
\newcommand{\pip}{\mbox{$\pi^{+}$}}
\newcommand{\pipm}{\mbox{$\pi^{\pm}$}}

\newcommand{\ks}{\mbox{$K_{S}$}}
\newcommand{\kl}{\mbox{$K_{L}$}}

\newcommand{\kpm}{\mbox{$K^{\pm}$}}

\newcommand{\kskl} {\mbox{\ks\kl}}
\newcommand{\kpkm} {\mbox{$K^{+} K^{-}$}}
\newcommand{\epem} {\mbox{$e^{+} e^{-}$}}

\newcommand{\fo}{\mbox{$f\!_{o}$}}
\newcommand{\ao}{\mbox{$a_{o}$}}

\newcommand{\reep}{\mbox{$\Re(\epsilon'/\epsilon)$}}

\newcommand{\kkbar}{\mbox{$K\overline{K}$}}

\newcommand{\klcha} {\mbox{$\kl\rightarrow \pip \pim$}}
\newcommand{\kscha} {\mbox{$\ks\rightarrow \pip \pim$}}
\newcommand{\klneu} {\mbox{$\kl\rightarrow \pio \pio$}}
\newcommand{\ksneu} {\mbox{$\kl\rightarrow \pio \pio$}}
\newcommand{\kpmd} {\mbox{$\kpm\rightarrow \pipm \pio$}}
\newcommand{\etapim} {\mbox{$N(\klcha) / N(\kscha)$}}
\newcommand{\etapio} {\mbox{$N(\klneu) / N(\ksneu)$}}
\newcommand{\Ratio}  {\mbox{$\etapim \over{\etapio}$}}
\newlength{\leftcolwidth}%
\newlength{\rightcolwidth}%
\newcommand{\doublecolumn}[3]%
{%
\setlength{\leftcolwidth}{#1}%
\setlength{\rightcolwidth}{\textwidth}%
\addtolength{\rightcolwidth}{-1.0\leftcolwidth}%
\addtolength{\rightcolwidth}{-1.5ex}%
\begin{minipage}{\leftcolwidth}%
{#2}
\end{minipage}%
\begin{minipage}{1.5ex}%
\mbox{\,}
\end{minipage}%
\begin{minipage}{\rightcolwidth}%
{#3}
\end{minipage}%
}

\newcommand{\AmS}{{\protect\the\textfont2
  A\kern-.1667em\lower.5ex\hbox{M}\kern-.125emS}}
\def\Title#1{\begin{center} {\Large {\bf #1} } \end{center}}

\begin{document}

\Title{A Status Report of KLOE at \dafne}

\bigskip\bigskip


\begin{raggedright}  

{\it Sergio Bertolucci\index{Bertolucci, S.}\\
I.N.F.N - Laboratori Nazionali di Frascati\\
Via E.Fermi 40, 00044 Frascati, Italy \\
for the KLOE collaboration}%
        \footnote{The KLOE collaboration: M.~Adinolfi, A.~Aloisio, F.~Ambrosino,
        A.~Andryakov, A.~Antonelli, M.~Antonelli, F.~Anulli, C.~Bacci,
        A.~Bankamp, G.~Barbiellini
% ...
Y.~Zhou.}
\bigskip\bigskip
\end{raggedright}


\begin{abstract}

The major goal of the  KLOE experiment is to measure  
$\Re(\epsilon' /\epsilon)$ in the $\kkbar$ system  with a precision of $10^{-4}$,
both via the traditional double ratio method and quantum 
interferometry. 
The experiment has started taking data at \dafne, the 
$\phi$-factory built at the Frascati National Laboratory (LNF) of
INFN in Italy, beginning in April 1999.
In the early phase of the commissioning (before KLOE roll-in), $\dafne$  has achieved 
multi bunch luminosities of $1 \times 10^{31} {\rm cm^{-2}\  s^{-1}}$ with 13 bunches and 200
mA.
Detailed studies are presently underway to compensate for the large perturbation 
brought in by the KLOE solenoid, which has caused a drastic decrease of the peak luminosity.
Nevertheless the first collected data, corresponding to an integral
luminosity of 220~${\rm nb^{-1}}$, show that KLOE is performing to the design
specifications in all its hardware and software components.     
\end{abstract}

\section{Introduction}

The Frascati $\phi$ factory, \dafne \cite{dafne}, is 
an \epem\ collider optimized to operate at center of 
mass energy of M$_{\phi}$ with very high luminosity 
(L$_{peak}=5 \times 10^{32}\ {\rm cm^{-2}\ s^{-1}}$).\\
Since the vector meson $\phi$ decays into \kpkm\ 49~\%\ , \kskl\  34~\%\ of 
the times, \dafne\ can be considered a  factory
of neutral and charged kaons, produced in collinear pairs, with momenta 
of $\sim$ 110 MeV/c and  in a pure quantum state ($J^{PC} = 1^{--})$.
Interference effects will then appear in the decays of the \kskl$($\kkbar$)$ pairs,
allowing the measurement of all (except one) CP and CPT violation parameters.\\ 
The observation of one charge conjugated state of the kaon guarantees {\it(tags)} 
the existence of the other one in the opposite direction.
A pure \ks\ beam {\em without using any regenerator}  is therefore available.\\
At the design luminosity, \dafne\ will produce, in one HEP physics year ($10^7$\ {\rm s}), 
$8.5 \times 10^9$ \kskl, $1.2 \times 10^{11}$ \kpkm, $2.5 \times 10^8$ $\eta$, 
$2.5 \times 10^6$ $\eta'$ and $\sim$  $10^5$ \ao\ and \fo\ . \\

A detector, KLOE, has been built to exploit the following main physics goals:
\begin{itemize}
\item the measurement of $CP/CPT$ parameters from interferometry and the double
ratio R, with a sensitivity of $10^{-4}$; \item the measurement of the kaon form factors, 
\ks\ rare decays and the \ks\ semileptonic asymmetry (not measured so far);

\item the study of  the radiative $\phi$ decays, investigating the nature 
of the \ao, \fo\ mesons  and providing a precise determination of the relative branching
ratio between $\phi \rightarrow \eta \gamma$ and $\phi \rightarrow 
\eta' \gamma$. 
\end{itemize}
With the first 100 pb$^{-1}$ of collected data we already expect to measure \reep\
with an error of $\sim$ 0.1\%, to improve the measurement of 
the kaon form factors and to carry out most of the program on the radiative $\phi$ decays.

Since the  double ratio (R) is related to \reep\ by:
\begin{equation}
  R  = \Ratio = 1+ 6 \reep
\end{equation}
\noindent
the final goal of the experiment of  measuring \reep\ with a precision of $10^{-4}$,
translates into a global error on R of  few $10^{-4}$.
While the production rate of the kaon pairs and the tagging efficiency for
the \ks\ and the \kl\ cancel out identically in the double ratio, this is not true for
the detection efficiencies, which for this reason have been kept as high as possible
in the detector design stage.\\
In addition, KLOE is a self-calibrating experiment: the abundance  of events such
as \kl\ $\rightarrow \pi^{+} \pi^{-} \pi^{0}$ ($K_{L}^{\pi\pi\pi}$) or \kpmd\,
which produce in the final state both charged and neutral pions, will allow 
the determination of the efficiencies from the data themselves.\\
Since at these energies the decay length ($\lambda$) of
the \ks\ and  \kl\ is  0.6~cm and 345~cm, event counting will have to be performed in two
different fiducial volumes. 
While  essentially all the \ks\  decay in vacuum
inside a beam pipe of radius 10~cm, which approximates an infinite decay volume,
the difference between the boundaries of 
charged and neutral decay volumes of the \kl\ has  to be known with a precision of 
$\sim$ 0.5 mm. Again, the use  of $K_{L}^{\pi\pi\pi}$ and \kpmd\  events will help in surveying and 
correcting for this difference.
 
Background subtraction is finally another source of systematics.  
Main backgrounds for the CP violating \kl\ decays are  $K_{L}\rightarrow \pi^0\pi^0\pi^0$ ($K_{L}^{000}$)
and $K_{L}\rightarrow \pi \mu \nu$ ($K_{\mu3}$) which have signal to noise
ratios of
1/240 and 1/135 respectively, so that  rejection factors of $8(6) \times 10^{-4}$ are
needed.

Statistics is the other half of the game: in order to reach the 
desired accuracy in \reep\  we need to collect 
$\sim 4 \times 10^6\  \klneu\ $, which can be done in two HEP years 
($2 \times10^{7}\ {\rm s}$) at  \dafne\ design luminosity $L_{0} \sim  5\times 10^{32}\ {\rm cm^{-2}
s^{-1}}$.\\ 

\section{\dafne\ and KLOE: the pilot run}
In order to reach this ambitious goal, \dafne\ has been designed to operate
at a conservative single bunch luminosity 
($L_{0} \sim L_{VEPP-2M} = 4 \times 10^{30} cm^{-2} s^{-1}$)
and with a large number of bunches (up to 120).
The \dafne\ complex consists of two coplanar rings, 
a Linac accelerating $e^{+}$ up to 550 MeV and $e^{-}$ to 800 MeV and 
a 510 MeV accumulator ring for fast topping-up.
The two beams ($\sigma_{x}, \sigma_{y}, \sigma_{z} = 20\ {\rm \mu m, 2\ mm, 3\ cm} $) 
intersect with
an half angle of $\sim$ 12 mrad to reduce parasitic interactions between ingoing and
outgoing bunches.
Crossing frequency is up to 368 MHz (with 120 bunches). Typical number of particles/bunch is 
$\sim 10^{11}$,
corresponding to $\sim$ 43 mA/bunch, with a total current per beam of 5.2 A.\\
A coupling  $\beta_{y} /\beta_{x}$  ($\sim$ 1\%)
with a  $\beta_{y}= 0.045\ {\rm m}$ , together with a rather aggressive value
for the tune-shift ($\xi  = 0.04$) have been chosen.
Before KLOE rolled-in, a single bunch luminosity $L \sim 1.5 \times 10^{29} cm^{-2} s^{-1}$ 
was achieved with 20 mA/beam and soon after  $ L \sim 10^{31} cm^{-2} s^{-1}$, with 13 bunches 
and $\sim$~200~mA/beam, was also reached.\\ 
KLOE insertion has introduced a large perturbation in the rings optics,
due to the large Bdl of the solenoid (2.4 T/m, to
be compared with a beam rigidity $ B\rho \simeq$ 1.7 T/m). If uncompensated, the KLOE field would
rotate the beam of $\sim 40$ degrees at the interaction point.\\
Compensation has to be obtained on both rings through the
careful tuning  of solenoidal compensators, while the quadrupoles of the low $\beta$
insertion need to be rotated, in order to reduce transverse coupling.
Only a first order pass has been made so far, which still cannot achieve neither
a satisfactory coupling. nor a good matching of the  $\beta^{*}$ on both rings.\\ 
At the moment of this conference, single bunch luminosities 
$L_{0} \sim  2\times 10^{29} cm^{-2} s^{-1}$ 
with 10 mA/bunch
and multi-bunch $L \sim 2\times 10^{30}\ {\rm cm^{-2} s^{-1}}$ with ten bunches have been
reached.

The KLOE detector was fully operational since July 1998 and a lot of
experience on the detector calibration, data acquisition and triggering 
has been gained running with cosmic rays.
The detector was installed in \dafne\ early in 1999 and was fully operational in March 1999. 
First collisions were observed and recorded on the 23rd April 1999.\\ 
Another few weeks were later dedicated to the study of the $\phi$ line shape
running in single bunch mode. Once the beam energy was 
determined, multi-bunch operation started, for a total of 10 days of stable operation.
The integrated luminosity of this pilot run
was $\sim 200 \hbox{ nb$^{-1}$}$, which allowed us to
tune-up the calibration and reconstruction  procedures.

\begin{figure}[htb]
\begin{center}
\epsfig{file=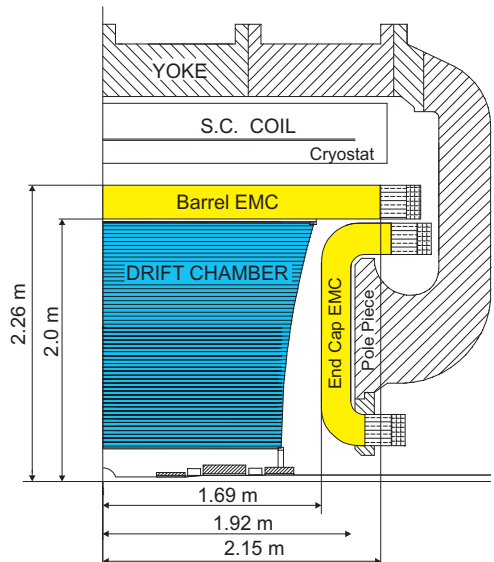,height=7.5cm}
\vspace{-0.5cm}
\caption{Schematic view of the KLOE detector}
\label{fig:kloe}
\end{center}
\end{figure}
\section{Overview of KLOE and its performances}
The KLOE detector \cite{kloe1} \cite{kloe2} is a general purpose 
\epem\ detector of respectable dimension ($\sim$ 7~m diameter, 6~ m length).
Going outward (fig. \ref{fig:kloe}),
a 0.5 mm thick cylindrical beryllium beam pipe with a spherical shape of
10~cm radius (16~$\lambda_{s}$) surrounds the interaction point.
The inner permanent quadrupole region is instrumented by two lead-scintillator
tiles calorimeters of $\sim$ 5 cm thickness, each organized in 16 wedges.
Their major function is to improve the rejection efficiency
for $K_{L}^{000}$.\\
Altough their reduced shower containement
results in a poor energy resolution,  $\gamma$ detection efficiency, timing and 
position resolution
($\epsilon \ge 98\%$, $\sigma_x \sim 5\ {\rm cm}$, $\sigma_t \sim 1\ {\rm ns}$) 
are enough to achieve the desired rejection power.\\
A large drift chamber, DCH \cite{kloe3}, of 4~m diameter and 3.5~m length, 
follows surrounded by
an  electromagnetic calorimeter, EMC, which covers almost  hermetically
the solid angle. 
Everything is embedded inside the coil cryostat and a specially shaped
iron yoke. The superconducting coil generates a field of 6 kG.
The trigger \cite{trig} and data acquisition, DAQ \cite{daq}, 
systems allow to collect a very high event rate.
At the highest luminosity we expect to have $\sim$ 2.5 KHz of $\phi$,
$\sim$ 50KHz Bhabha, 2.5 KHz cosmics and few KHz of machine
background.
DAQ is able to handle a data throughput well above 50 MB/sec, 
through two levels of data  concentration via dedicated hardware processors 
and online event building in parallel farms.
The two levels trigger is based on  EMC energy deposit and DCH hit multiplicity. 
Its main task is to reduce the
event rate to the one allowed by DAQ without loosing any relevant 
$\phi$ decay, while retaining some portions of Bhabha and cosmics for
calibration purposes.\\
The main task of the EMC is to reconstruct the $K_{L}^{00}$ decay vertex 
and to efficiently reject the $K_{L}^{000}$ background.
The vertex reconstruction of the decay 
$K_{L}\rightarrow \pi^0\pi^0$ is performed by accurately measuring 
the arrival time to the calorimeter
of all the photons \cite{kloe1,kloe2}. Using the flight direction of the
$K_{S}\rightarrow \pip\pim$, one can easily demonstrate that a single photon is sufficient
to determine the \kl\ decay vertex. 
These requirements correspond to the following design specifications:

(1) $\sigma_{E}/E$  $\sim$  $5\% / \sqrt{E\hbox{ (GeV)}}$;
(2) $\sigma_{T} \simeq 70 \hbox{ ps} / \sqrt{E\hbox{ (GeV)}}$;

(3) full efficiency for $\gamma$'s in the range 20--280 MeV;
(4) hermeticity.\\
The KLOE EmC is a fine sampling lead-scintillating fiber calorimeter with 
photomultiplier (PM) read-out. The central part (barrel) approximating 
a cylindrical shell 
of 4 m inner diameter, 4.3 m active length and 23 cm thickness 
($\sim 15$ $X_0$), consists of 24 modules.
Two end-caps, consisting of 32 ``C'' shaped modules, close hermetically 
the barrel. 
The modules are read out on the two sides with a granularity 
of$\sim 4.4 \times 4.4$ cm$^2$  by fine mesh PM's,
for a  total number of $4880$ channels.
The basic calorimeter structure consists in 
an alternating stack of 1 mm scintillating fiber layers glued between 
thin grooved lead foils. 
The final composite has a fiber:lead:glue volume 
ratio of approximately 48:42:10,
a density of $\sim 5$ g/cm$^3$ and a $X_0$ of $\sim 1.6$ cm.

\begin{figure}[htb]
{\doublecolumn
{0.5\textwidth}
{
\vspace{1.8cm}
\begin{center}
\epsfig{file=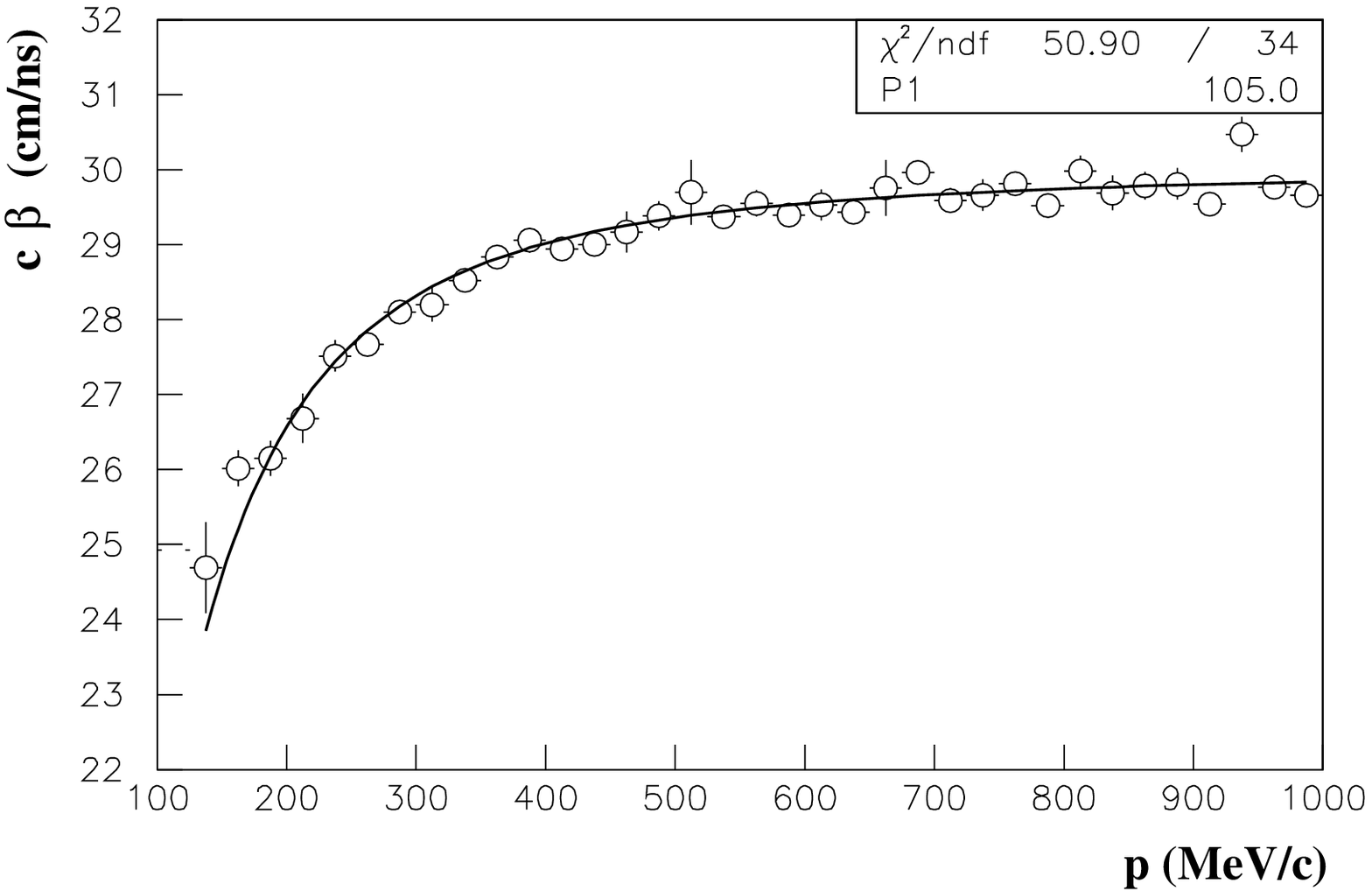,width=8.1cm}
\end{center}
}
{
\begin{center} 
\epsfig{file=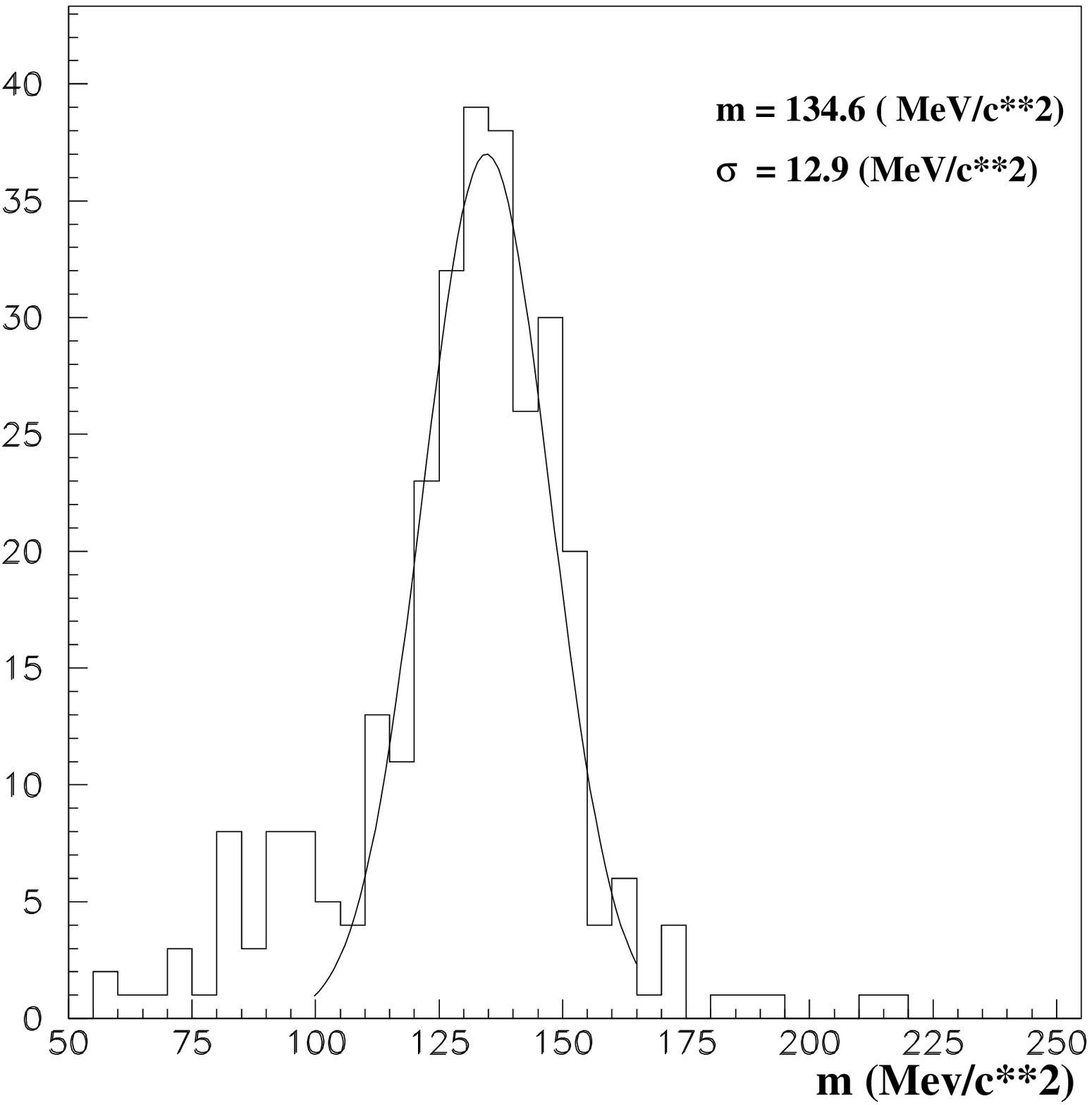,width=7.1cm}
\end{center}
}
}
\caption{EMC performances left): $c \beta $ vs P(Mev/c) for cosmic ray events,
the fit is described in the text; right): Invariant mass of the two  $\gamma $'s coming from $\pi^{0}$ in 
$\pi^{+} \pi^{-} \pi^{0}$ events. }
\label{fig:emc}
\end{figure}

The EmC modules and the front-end electronics were 
fully installed in KLOE at the beginning of 1998. 
The calorimeter has been fully operational since then.
After the installation, a first calibration of 
the calorimeter with cosmic rays has been performed, to obtain
the minimum ionizing peak for each calorimeter channels with
an accuracy better than 1\%.
The responses of all PMs is equalized to a few percent, in order not to 
bias the trigger response.
The times measured at the two ends of a cell allow to obtain the
arrival time of the particles and the coordinate along the fiber.\\
Fitting the measured time pattern for the
fired cells in high energy cosmic ray events, the time offset of each
channel is determined with a precision of $\sim$ 10~ps.
A check of this calibration consists in measuring the velocity
of cosmic rays  as a function of their momentum, as reconstructed by the drift chamber.
A fit to the measured distribution of $\beta = p / \sqrt{p^2+m^2}$, leaving the mass $m$ 
as the only free parameter, yields a value
in good agreement with the muon mass, as shown in fig. \ref{fig:emc}-left.

The Bhabha and  $e^+e^-\rightarrow \gamma \gamma$  events allow then to set
the energy and time scales.
The measured energy resolution at 510 MeV is $\sim$ 7.8~\% ,
corresponding to $\sigma/E \simeq 5.5 \% / \sqrt{E(\hbox{GeV})}$.
As a check of the energy scale calibration, we reconstructed the 
invariant mass of the
photon pair from $\phi\rightarrow \pi^+\pi^-\pi^0$ events.
The peak in the invariant mass distribution is within 1 MeV from the 
$\pi^0$ mass, 
as shown in fig. \ref{fig:emc}-right. 

In order to evaluate the time performances of the calorimeter, 
the quantity $\Delta t =  t_{clu1}-R_{1}/c-(t_{clu2}-R_{2}/c)$ 
is measured for the two photons of
$e^+e^-\rightarrow \gamma\gamma$ events; $t_{clu}$ is the arrival 
time of the particle  as measured by the calorimeter cluster,
$R$ is the position of the cluster centroid.
We observe $\sigma_{\Delta T} \sim  150\ $ps, corresponding to 
$\sigma_t \sim $ $75 \hbox{ ps} / \sqrt{E\hbox{ (GeV)}}$;

The DCH should provide 3-D tracking with a resolution of 
$\sim$ 200 $ {\rm \mu m}$ in the bending plane and a z-resolution
of $\sim$\  1 mm on the decay vertices over the whole sensitive volume.
In order to efficiently reject the $K_{\mu 3}$ background, a momentum resolution of 
0.5\% for low momentum tracks is required; the active volume should be as 
transparent as possible to minimize multiple scattering.
Since the \kl's decay almost uniformly all over the whole chamber, the 
tracking reconstruction efficiency should have minimal dependence on the 
position in the DCH.
The design of the tracker is also driven by the 
requirement of having a light and homogeneous active volume to 
prevent \kl\ regeneration and photon conversions; for the same reason
low-mass walls are necessary.
The adopted solution is a cylindrical drift chamber whose supporting structure 
is entirely made of carbon-fiber and filled with an ultra-light gas mixture 
(90\% He-10\% iC$_{4}$H$_{10}$).
The requirement of a 3-D reconstruction and of uniform tracking
forces the choice of square cells arranged in 
layers with alternating stereo angles.
In order to keep a constant
drop, the stereo angle varies with increasing radius from $\pm$60 mrad
to $\pm$ 150 mrad. The cells are organized in 12 inner layers
of smaller cells ($2\times 2\ {\rm cm^2}$) and 46 outer layers of bigger
cells ($3\times 3\ {\rm cm^2}$).
\begin{figure}[htb]
{\doublecolumn
{0.5\textwidth}
{
\begin{center}
\epsfig{file=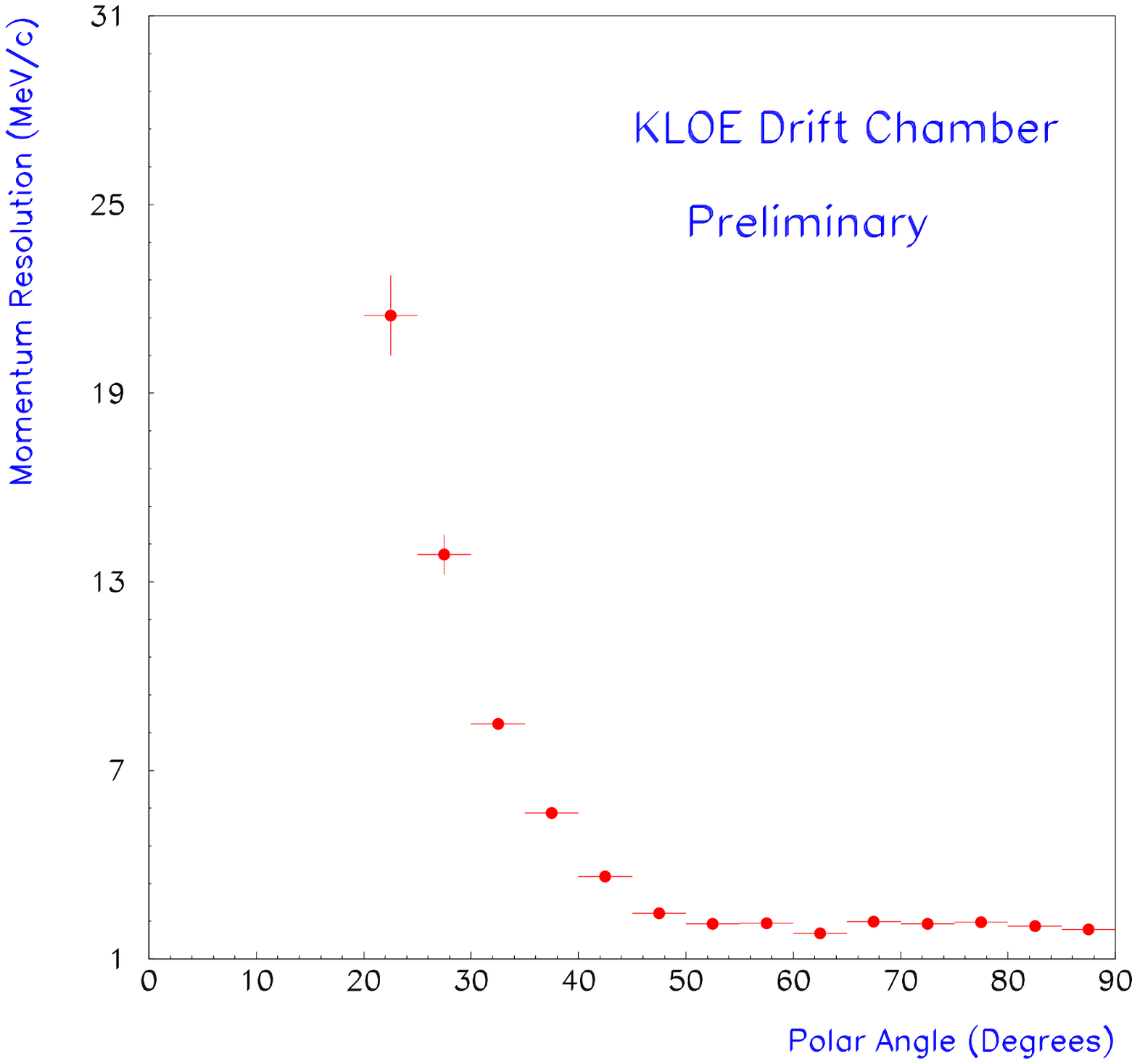,width=7.1cm}
\end{center}
}
{
\begin{center} 
\epsfig{file=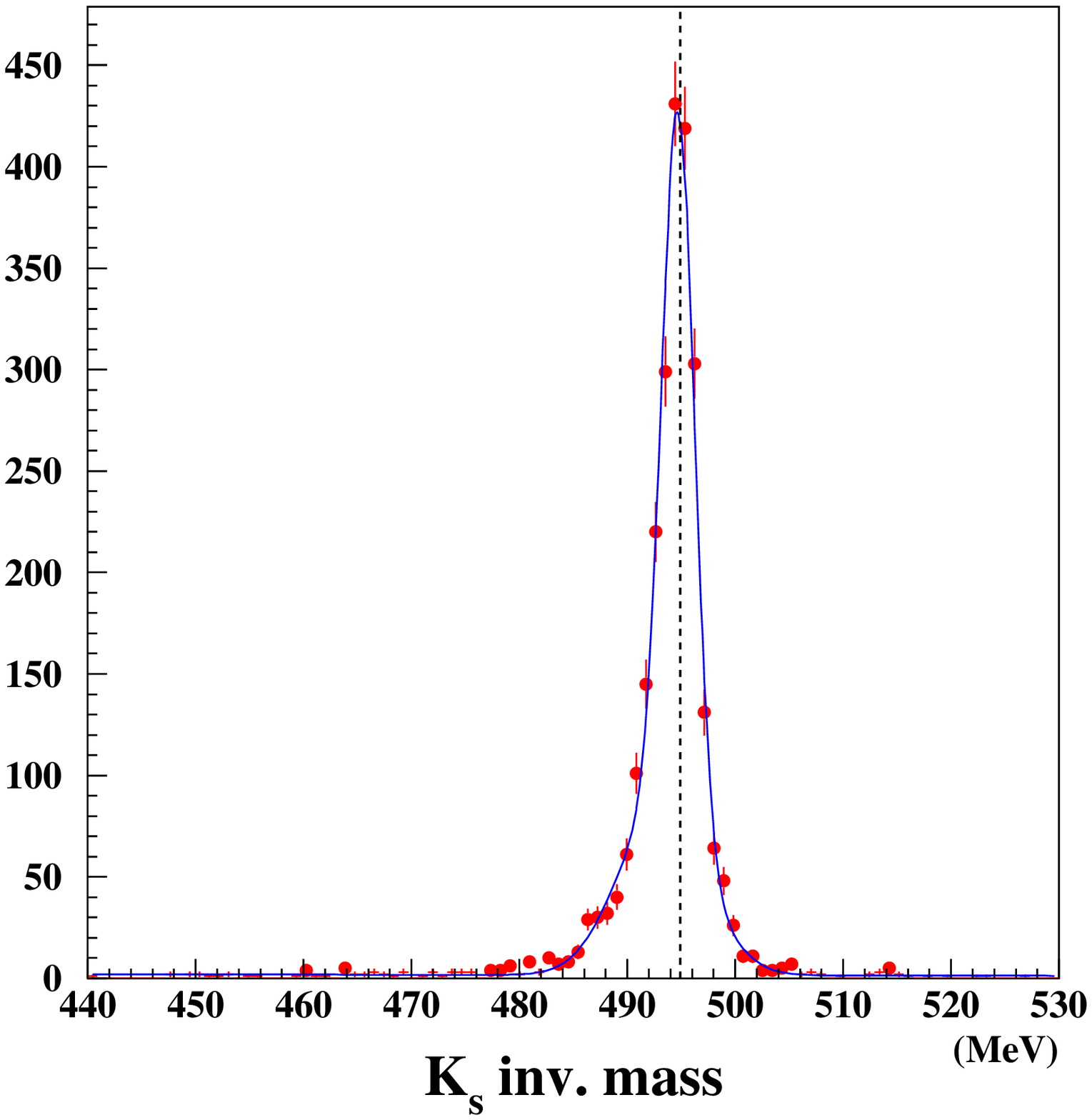,width=7.1cm}
\end{center}
}
}
\caption{DCH performances: left): momentum resolution as function of $\theta$ angle for Bhabha events,
right): $\pi^{+} \pi^{-}$ invariant mass for  \ks\ decays. }
\label{fig:dch}
\end{figure}
Good efficiency and space resolution, together with negligible ageing
effects, are obtained running at a gas gain of $\sim$
10$^{5}$. This requirement is met with 25~${\rm \mu m}$ W(Au) sense wires
and 80~${\rm \mu m}$ field Al(Ag) wires kept at  $\sim$ 1.9 KV. 
The DCH has a total of 52000 wires  with a ratio 3:1 between the number 
of field to sense wires.

The DCH was moved inside KLOE on April 1998 and
has been kept in operation for more than 1 year with 
a very low number of dead/hot channels/wires (below 0.1\%).
The calibration of the drift chamber proceeds, after subtracting the time
offsets of each single wire, via an iterative procedure which minimizes the
fit residuals  by redefining the $s-t$ relations.
In this  gas mixture we do not expect to have a saturated drift velocity.
The $s-t$ relations are therefore not linear and they are parametrized
with a 5th order Chebychev polynomial.
The minimization procedure stops
whenever the mean of the fit residuals is below 100 $\mu m$.
The behaviour of the space resolution
along the cell shows the usual dependence on primary ionization and
longitudinal diffusion. The average value of the resolution is
$\sim$ 150 $\mu m$.
The whole chamber calibration can be performed in $\sim$ 4 hours, using cosmic rays. 

Using Bhabha events and $K_{s} \rightarrow \pi^{+} \pi^{-} $,
we can then evaluate the momentum resolution and the momentum scale.
As shown in fig. \ref{fig:dch}-left, we obtain a momentum resolution 
better than 0.4\%
for polar angles greater than 45$^{\circ}$.
The invariant mass of the \ks\ is also shown in fig. \ref{fig:dch}-right; 
fitting it
with a gaussian, we obtain M$_{K_s} \sim 496$ MeV with $\sigma$ of $\sim$ 1 MeV.

\section{CONCLUSIONS}
KLOE has started operations on the beam.\\
$\dafne$ is presently
delivering a luminosity L of $\sim 2 \times 10^{30} {\rm cm^{-2} s^{-1}}$.\\
Using the 200 ${\rm nb^{-1}}$ of delivered luminosity so far, the detector has shown to be 
fully operational in all its hardware and software components, reaching or being very
close already to its design specifications.
KLOE is ready for real data.

\end{document}